\documentclass[prl, showpacs, subeqn, graphicx]{revtex4}
\usepackage{amsmath}
\usepackage{graphicx}
\usepackage[dvips]{color}

\begin{document}

\title{Optoacoustic solitons in Bragg gratings}
\author{Richard S. Tasgal and Y. B. Band}
\affiliation{Departments of Chemistry and Electro-Optics, 
Ben-Gurion University of the Negev, Beer-Sheva 84105, Israel}
\author{Boris A. Malomed}
\affiliation{Department of Interdisciplinary Studies, School of 
Electrical Engineering, Faculty of Engineering, Tel Aviv 
University, Tel Aviv 69978, Israel}

\begin{abstract}
Optical gap solitons, which exist due to a balance of nonlinearity and
dispersion due to a Bragg grating, can couple to acoustic waves
through electrostriction.  This gives rise to a new species of
``gap-acoustic'' solitons (GASs), for which we find exact analytic
solutions.  The GAS consists of an optical pulse similar to the
optical gap soliton, dressed by an accompanying phonon pulse.  Close
to the speed of sound, the phonon component is large.  In subsonic
(supersonic) solitons, the phonon pulse is a positive (negative)
density variation.  Coupling to the acoustic field damps the solitons'
oscillatory instability, and gives rise to a distinct instability for
supersonic solitons, which may make the GAS decelerate and change
direction, ultimately making the soliton subsonic.
\end{abstract}

\pacs{42.81.Dp, 42.70.Qs, 43.25.+y, 72.50.+b}

\date{\today}

\maketitle

\textit{Introduction.}---Light and sound tend to have widely disparate
frequencies, wavelengths, and velocities.  As a consequence,
interaction between optical and acoustic signals is often weak.  The
interaction is important, e.g., for Brillouin scattering, diffraction
of light by ultrasonic waves \cite{Boyd.2003}.  Brillouin scattering
in optical fibers---coupling of longitudinal or transverse acoustic
modes to optical waves through electrostriction---has been analyzed
extensively, though generally not for solitons
\cite{ThomasRowellDrielStegeman.1979, Shelby, BucklandBoyd, Fellegara,
GranotSternklar.BrillouinFilter}.  Sound waves acting on trains of
optical solitons in fibers have been studied
\cite{SmithMollenauer.1989, DianovLuchnikov}, but these interactions
were essentially between acoustic continuous waves and their optical
counterparts, subject to periodic modulation.  Brillouin scattering
has been employed in acousto-optic filters, in which light diffracts
off an acoustic grating \cite{GranotSternklar.BrillouinFilter}.
Brillouin scattering in fibers can lead to pulse compression
\cite{Hon.1980} or formation of trains of very narrow optical solitons
\cite{BongrandMontes.2001}.  Interaction of phonons with optical
nonlinear Schr\"{o}dinger solitons, with nonlinearity due to
electromagnetically-induced transparency, was analyzed in
Ref.~\cite{Sazonov}.  Reference \cite{Zabolotskii.2004} considered
solitons in Bragg gratings, with electrostriction the only
nonlinearity; the correct soliton solution is a limiting case of the
results below.

In this work, we consider light and acoustic waves in a cubic
($\chi^{(3)}$) nonlinear medium with a Bragg grating, a fixed
spatially periodic variation in the index of refraction
\cite{SterkeSipe.1994}.  Bragg gratings are typically written in the
cladding of an optical fiber, though they have been induced by
copropagating continuous waves \cite{SimaeysCoen.2004}.  Fiber Bragg
gratings without electrostriction are known to support optical gap
solitons, nonlinear solitary waves with frequencies in the band gap
created by the Bragg grating \cite{ChenMills.1987, SterkeSipe.1994,
experiment}.  Gap solitons can have arbitrarily small velocity, which
has motivated great interest in them for stopping light, though
creating very slow gap solitons experimentally has been difficult.
Early experimental realizations of gap solitons had velocities
approximately half the speed of light \cite{experiment}.  Various
methods have been proposed for creating slow gap solitons: building up
slow gap solitons directly by Raman amplification
\cite{WinfulPerlin.2000}, obtaining quiescent solitons from collisions
of fast-moving gap solitons \cite{MakMalomedChu.2003}, and retarding
fast solitons by passing them through gradually varying (``apodized'')
Bragg gratings \cite{MakMalomedChu.2004}.  In a recent experimental
breakthrough, optical gap solitons were slowed to a sixth the speed of
light using apodized Bragg gratings \cite{MokSterke.2006}.

Electrostriction, compression of the medium due to variations of the
intensity of light, allows light to drive acoustic waves; conversely,
acoustic waves feed back into the optical field through the dependence
of the refractive index on the material density.  Gap solitons can be
arbitrarily slow; at velocities close to the speed of sound, an
\emph{opto-acoustic resonance} may dramatically enhance the interplay
between light and sound.  We derive a generalization of the standard
optical gap soliton, including an acoustic component.  The dynamics of
these ``gap-acoustic solitons'' (GASs) are significantly different
from optical gap solitons without the acoustic dressing when the
velocities are slow, and slightly different when velocities are large.
Our results demonstrate that if an optical gap soliton is slowed to a
velocity on the order of the speed of sound, the inherent
electrostrictive effects can do the work of slowing the soliton to
significantly below the speed of sound, even creating stopped light.
The results also suggest that GASs may be controlled by continuous
acoustic waves or pulses.

We assume one transverse electromagnetic mode and one acoustic mode,
either because the waveguide has exactly one transverse optical mode
and one acoustical mode, or because coupling to other modes is
negligible.  We have derived effectively 1+1-dimensional coupled mode
equations using standard methods \cite{SterkeSipe.1994, Agrawal.1989}
for light in a waveguide, adding a phonon field and acousto-optic
coupling induced by electrostriction \cite{Boyd.2003,
ThomasRowellDrielStegeman.1979, Hon.1980, Shelby,
SmithMollenauer.1989, DianovLuchnikov, BucklandBoyd, Fellegara}:
\begin{subequations}
\label{governing_equations}
\begin{eqnarray}
0 & = & i k_0' u_t
      + i u_z
      + \kappa v
      + \frac{2\pi (\omega_0/c)^2}{k_0 A}
	\left(
	  \chi_\mathrm{s} |u|^2
	+ \chi_\mathrm{x} |v|^2
	\right) u
      + \chi_\mathrm{es} \, w \, u \, , \\
0 & = & i k_0' v_t
      - i v_z
      + \kappa^* u
      + \frac{2\pi (\omega_0/c)^2}{k_0 A}
	\left(
	  \chi_\mathrm{x} |u|^2
	+ \chi_\mathrm{s} |v|^2
	\right) v
      + \chi_\mathrm{es} \, w \, v \, , \\
0 & = & w_{tt}
      - \Gamma \, w_{tzz}
      - \beta_\mathrm{s}^2 \, w_{zz}
      + \lambda (|u|^2 + |v|^2)_{zz} \,.
\end{eqnarray}
\end{subequations}
Here $z$ is the coordinate along the waveguide and $t$ is time;
$u(z,t)$ and $v(z,t)$ are slowly varying envelopes of the
counterpropagating electromagnetic waves \{total electric field
$E(z,t) = u(z,t) \exp[i(k_0 z - \omega_0 t)] + v(z,t) \exp[-i(k_0 z +
\omega_0 t )] + \mathrm{c.c.}$\}, with carrier frequency $\omega_0$
and wave numbers $\pm k(\omega_0) \equiv \pm n(\omega_0)\omega_0/c$,
$n(\omega)$ being the index of refraction, $k_0' \equiv
dk/d\omega|_{\omega=\omega_0}$ the reciprocal group velocity, and
$\kappa$ the Bragg reflectivity.  The self- and cross-phase modulation
coefficients are $\chi_\mathrm{s}$ and $\chi_\mathrm{x}$, and $A$ is
the effective cross-section area of the waveguide modes.  The acoustic
field $w(z,t)$ is the density of the medium, offset by a constant
$W_0$.  Equations~(\ref{governing_equations}) assume that the acoustic
wavelengths are much larger than the optical one, so sound waves are
not subject to Bragg reflection.  The form of the electrostrictive
interaction involves the average square electric field $\langle E^2
\rangle = 2(|u|^2+|v|^2)$ (see Eqs.~(9.3.11-12) in
Ref.~\cite{Boyd.2003}).  The electrostrictive coefficient $\lambda$,
which comes from the dependence of the energy density on the index of
refraction $n(\omega,W_0+w)$, is $\lambda \equiv (2\pi)^{-1} (W_0+w)
\, n \, dn/dw$ \cite{Boyd.2003}; in a waveguide, the coefficient may
be strongly affected by the modal structure
\cite{ThomasRowellDrielStegeman.1979, BucklandBoyd, Fellegara}.  The
other electrostrictive coefficient, $\chi_\mathrm{es} = (\omega_0/c)
\, dn/dw$, comes from local changes in the wave vector [$k =
n(\omega_0, W_0+w) \omega/c$], which depends on the index of
refraction, which in turn depends on the acoustic field
\cite{Boyd.2003, Agrawal.1989, DianovLuchnikov, BucklandBoyd}.  The
speed is sound is denoted by $\beta_s$.  Phonon viscosity $\Gamma$ is
included for completeness, though current Bragg waveguides are not
long enough for it to be significant.  Mass
\begin{equation*}
  M = A \int_{-\infty }^{\infty} w(z,t) dz
\end{equation*}
and number of photons
\begin{equation*}
  N = \frac{n(\omega_0)^2}{4\pi \hbar\omega_0} A
      \int_{-\infty}^{\infty} \left( |u|^2+|v|^2 \right) dz
\end{equation*}
are conserved; momentum
\begin{equation*}
  P = \frac{n(\omega_0)^2}{4\pi\omega_0}
       A \int_{-\infty}^{\infty}
       \left[
         \frac{i}{2} ( u \, u^*_z - u^* u_z
                     + v \, v^*_z - v^* v_z )
       - \frac{\chi_\mathrm{es}}{k_0' \lambda}
         r_z r_t
       \right]
       dz \, ,
\end{equation*}
where $r(z,t) \equiv \int^z w(z',t) dz'$ is an acoustic potential, is
invariant for $\Gamma=0$, but decays when viscosity is non-zero.

Equations~(\ref{governing_equations}) have resemblances to the
Zakharov system \cite{Zakharov_system, Zakharov_applications}: both
contain dispersive high-frequency waves, coupled to nondispersive
acoustic modes. The dissimilarity is the source of dispersion---a
Bragg grating rather than intrinsic material dispersion.  Looking at
Eqs.~(\ref{governing_equations}) as a Zakharov system with a Bragg
grating helps to identify realizations in additional physical systems:
surface and bulk acoustic waves in solids, where the grating (for the
surface mode) is a surface superstructure; coupled vibron and acoustic
waves in polymer molecules, with the grating (for the vibrons)
provided by an intramolecular structure; and coupled Langmuir and
ion-acoustic waves in plasmas (the original physical Zakharov system),
with a grating in the form of a dust crystal \cite{dust}.

\begin{figure}[tbp]
\centering
\includegraphics[width=8.6cm]{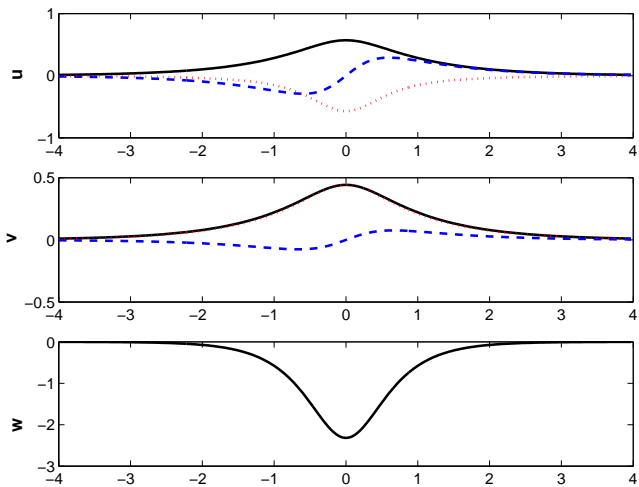}
\caption{(Color online) Three components of a supersonic gap-acoustic
soliton (right- and left-traveling envelopes of the electromagnetic
waves and the acoustic field), for $k_0' = \kappa = 2\pi
(\omega_0/c)^2 (k_0 A)^{-1} \chi_\mathrm{x} = \chi_\mathrm{es} = 1$,
$\chi_\mathrm{s} = \chi_\mathrm{x} / 2$, $\beta=0.25$, $\beta_s=0.2$.
$\Gamma=0$, and $\lambda=0.1$. Dashed, dotted, and solid lines show,
respectively, the real and imaginary parts and absolute values of the
fields.}
\label{GAS_fig1}
\end{figure}

\textit{Soliton solutions.}---We found a family of GAS solutions to
Eqs.~(\ref{governing_equations}) with zero phonon viscosity
($\Gamma=0$),
\begin{subequations}
\label{soliton:formulas}
\begin{eqnarray}
u & = & \sqrt{\kappa \gamma (1 + \beta \, k_0')} \, \alpha \,
	\sin Q \;
	\mathrm{sech}(\zeta \sin Q - i Q / 2)
	\exp [i \, \theta(\zeta) - i \, \tau \cos Q] \, , \\
v & = & -\sqrt{\kappa^* \gamma (1-\beta \, k_0')} \, \alpha \,
	\sin Q \;
	\mathrm{sech}(\zeta \sin Q + i Q / 2)
	\exp [i \, \theta(\zeta) - i \, \tau \cos Q] \, , \\
w & = & \frac{\lambda |\kappa \, \alpha^2|}
	     {\beta_s^2 - \beta^2}
	\frac{4\gamma \sin^2 Q}
	     {\cosh (2\zeta \sin Q) + \cos Q} \, ,
\end{eqnarray}
\end{subequations}
with
$\tau \equiv \gamma |\kappa| [t/k_0' - (\beta \, k_0') z]$,
$\zeta \equiv \gamma |\kappa| (z - \beta \, t)$,
$\gamma \equiv [1 - (\beta \, k_0')^2]^{-1/2}$,
and
\begin{subequations}
\label{soliton:coefficients}
\begin{eqnarray}
\theta(\zeta)
& \equiv & \frac{4 |\alpha|^2 (\beta \, k_0')}{1 - (\beta \, k_0')^2}
           \left(
             \frac{2\pi(\omega_0/c)^2}{k_0 A}
             \chi_\mathrm{s}
           + \frac{\chi_\mathrm{es} \lambda}
                  {\beta_s^2 - \beta^2}
           \right)
           \tan^{-1}\! [\tanh (\zeta \sin Q) \tan (Q/2)]
           \, ,
           \label{soliton:coefficient:theta}
\\
|\alpha|^{-2}
& \equiv & \frac{2\pi (\omega_0/c)^2}{k_0 A}
           \left(
             \chi_\mathrm{x}
           + \chi_\mathrm{s}
             \frac{1 + (\beta \, k_0')^2}{1 - (\beta \, k_0')^2}
           \right)
         + \frac{2 \chi_\mathrm{es} \lambda}
                {(\beta_s^2 - \beta^2)[1 - (\beta \, k_0')^2]}
         \, .
         \label{soliton:coefficient:alpha}
\end{eqnarray}
\end{subequations}
The solitons are characterized by two parameters: $\beta$, the soliton
velocity, and $Q$, which takes values $0 < Q < \pi$ and determines the
soliton's full width at half-maximum, $(|\kappa| \gamma \sin Q)^{-1}
\cosh^{-1}(2+\cos Q)$, peak intensity, and frequency [in the rest
frame, $\gamma |\kappa| (k_0')^{-1} \cos Q$].  These solutions also
hold for quiescent solitons ($\beta=0$) with non-zero phonon viscosity
($\Gamma>0$).  The velocity may take any value up to the group
velocity of light, $|\beta| < 1/k_0'$, except for a
\emph{velocity-spectrum gap} at and above the speed of sound, $|\beta|
\not\in [\beta_s, \beta_\mathrm{cr}]$, with
\begin{equation}
\beta_\mathrm{cr}^2 = \frac{1}{2 (k_0')^2}
		      \frac{\chi_\mathrm{x} + \chi_\mathrm{s}}
			   {\chi_\mathrm{x} - \chi_\mathrm{s}}
		    + \frac{\beta_s^2}{2}
		    \pm
			\sqrt{
			\left(
			  \frac{1}{2 (k_0')^2}
			  \frac{\chi_\mathrm{x} + \chi_\mathrm{s}}
			       {\chi_\mathrm{x} - \chi_\mathrm{s}}
			- \frac{\beta_s^2}{2}
			\right)^2
			- \frac{k_0 A}{2\pi (\omega_0/c)^2}
			  \frac{2 \chi_\mathrm{es} \lambda / (k_0')^2}
			       {\chi_\mathrm{x} - \chi_\mathrm{s}}
			} \, .
\end{equation}
For the typical case, $\chi_\mathrm{x} = 2 \chi_\mathrm{s} > 0$, the
negative sign applies.  For subsonic ($|\beta| < \beta_s$) solitons
[supersonic solitons ($\beta_s < \beta_\mathrm{cr} < |\beta| <
1/k_0'$)] , the density perturbation around the soliton is positive
[negative], i.e., compression [rarefaction].  The light components
($u,v$) vanish as the velocity approaches the speed of sound from
below, $|\beta| \rightarrow \beta_s$; both light and sound ($w$)
amplitudes diverge when the velocity approaches $\beta_\mathrm{cr}$
from above.  Figure~\ref{GAS_fig1} shows a moderately supersonic
soliton, with frequency in the middle of the bandgap ($Q=\pi/2$), and
velocity $\beta = (1.25)\beta_s = 0.25$.

With electrostriction, $\lambda \neq 0$,
Eqs.~(\ref{governing_equations}) do not admit solitons without an
acoustic component, but purely acoustic waves (solutions to the free
D'Alembert equation for $w$) are possible.
Equations~(\ref{governing_equations}) do not admit stable breathers,
as oscillations of a localized mode will generate emission of acoustic
waves, dissipating the oscillations.

\textit{Stability.}---We tested soliton stability by numerical
simulation of Eqs.~(\ref{governing_equations}), using a split-step
fast Fourier transform scheme \cite{Agrawal.1989}.  Simulations were
carried out systematically for three values of the soliton
coefficient, $Q=\pi/3$ (in the middle of the top half of the band
gap), which, for gap solitons without electrostriction ($\lambda=0$)
\cite{ChenMills.1987}, is well inside the stable region; $Q=\pi/2$
(mid-point of the band gap), which is stable but close to the
instability border; and $Q=2\pi/3$ (in the middle of the bottom half
of the band gap), which has an oscillatory instability
\cite{gap_soliton_instability}. We took ten values of the
electrostrictive coefficient $\lambda$, ranging over four orders of
magnitude, and the limit with zero instantaneous Kerr nonlinearity
$\chi_\mathrm{s} = \chi_\mathrm{x} = 0$.  The self- to cross-phase
ratio was $\chi_\mathrm{s}/\chi_\mathrm{x} = 1/2$.  The speed of sound
was taken to be $\beta_s = 0.2/k_0'$, which is faster than physically
realistic, but more clearly illustrates the dynamics.  Initial soliton
velocities were taken over a range from zero to twice the speed of
sound.  To impose consistent perturbations, all numerical simulations
had initial light amplitudes 1\% greater than those of the exact
soliton solutions.

\begin{figure}[tbp]
\centering
\includegraphics[width=8.6cm]{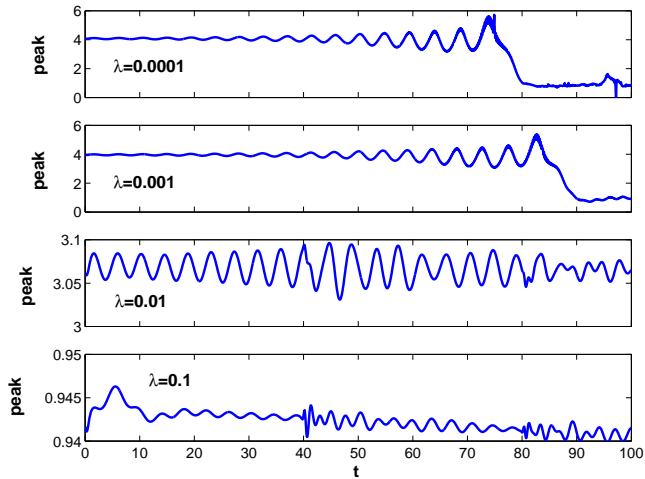}
\caption{(Color online) Time evolution of the peak power of the GAS,
[$\max_{z}(|u(z,t)|^2 + |v(z,t)|^2)$], for electrostrictive
coefficients $\lambda=0.0001$, $0.001$, $0.01$, and $0.1$.  The
solitons are taken with zero velocity and intrinsic parameter
$Q=2\,\pi/3$, which is unstable in the standard gap soliton model
($\lambda=0$).  The speed of sound is $\beta_s=0.2$.}
\label{GAS_fig2}
\end{figure}

Like gap solitons without electrostriction
\cite{gap_soliton_instability}, GASs are subject to oscillatory
instabilities, which grow until they destroy the solitons.  However,
electrostriction decreases the growth rate of the instability, as
shown in Fig.~\ref{GAS_fig2}.  Similarly, numerical simulations showed
that the growth rate of the oscillatory instability decreases for
velocities closer to the speed of sound.  A general trend is that
larger phonon components produce greater damping of the oscillatory
instability.  The damping may be due to emission of acoustic waves,
and that the GAS's acoustic component creates an effective trapping
potential which suppresses disintegration of the optical components.

\begin{figure}[tbp]
\centering\includegraphics[width=8.6cm]{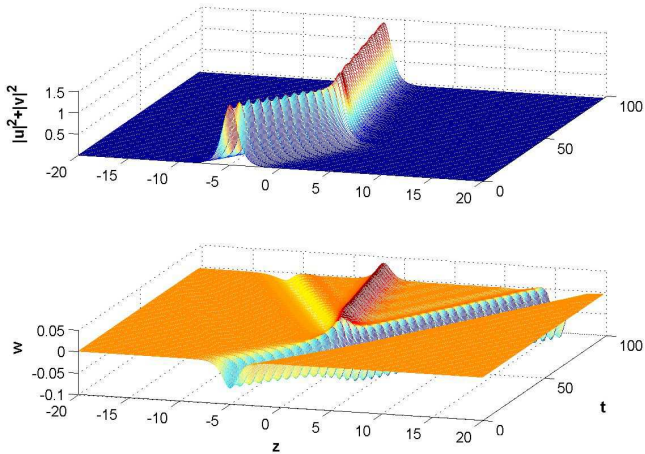}
\caption{(Color online) Instability of a supersonic GAS, for $Q=\pi/3$
and initial velocity $\beta=0.25$, which is $125$\% the speed of sound
$\beta_s=0.2$.  The electrostrictive coefficient is $\lambda=0.001$.}
\label{GAS_fig3}
\end{figure}

All solitons that are stable without electrostriction remained stable
with electrostiction $\lambda \neq 0$, if the velocity was subsonic
$|\beta| < \beta_s$.  All supersonic ($\beta_s < \beta_\mathrm{cr} <
|\beta| < 1/k_0'$) GASs were found to be unstable if numerically
integrated for long enough.  This supersonic instability is
qualitatively different from the oscillatory instability, and is
unknown for optical gap solitons without electrostriction.  The
supersonic instability may be connected to the fact that for slightly
supersonic solitons, the soliton momentum decreases with velocity
above the speed of sound $\frac{dP}{d\beta} < 0$, which is a known
instability criterion \cite{Zakharov_applications}.  Supersonically
unstable GASs tend to retain their integrity through growth of the
instability.  The closer is the supersonic soliton's velocity to the
speed of sound, the sooner the instability takes effect; this may be
explained by the divergence of the momentum slope $\frac{dP}{d\beta}$
at the critical velocity.  The changes in velocity are fairly abrupt,
and are accompanied by emission of phonons, which carry off momentum.
The solitons may change speed and direction a few times before
eventually settling to a stable subsonic form.  When the
electrostrictive coefficient is larger, and the phonon field larger,
the supersonic instability tends to be stronger: it occurs sooner, and
the soliton is more likely to be destroyed.  Destruction seems to be
associated with the decelerating GAS spending significant time in the
forbidden velocity region ($\beta_s \leq |\beta| <
\beta_\mathrm{cr}$).  Figure~\ref{GAS_fig3} shows an unstable
supersonic GAS, in which the soliton retains its integrity, settling
to a stable subsonic GAS.

\textit{Summary.}---Electrostriction couples acoustic waves to light
in a waveguide with a Bragg grating, and the resulting system supports
``gap-acoustic'' solitons, a generalization of optical gap solitons.
Electrostriction damps the gap soliton's (known) oscillatory
instability, and gives rise to another instability, which occurs when
the soliton's velocity exceeds the speed of sound.  In contrast to the
oscillatory instability, which destroys the gap soliton, the
supersonic instability tends to leave the soliton intact, transforming
it into a stable subsonic soliton.  Thus, electrostriction adds a new
level of complexity to the structure of optical gap solitons, and may
facilitate the creation of ultra-slow optical pulses.
Electrostriction and acoustic waves are present in virtually all
materials, so an understanding of physically realistic optical gap
solitons, especially slow ones, must entail a grasp of the effects of
electrostriction.

\begin{acknowledgments}
This work was supported, in part, by the U.S.-Israel Binational
Science Foundation (grant No.~2002147), the Israel Science Foundation
through the Center-of-Excellence grant No.~8006/03, the German Federal
Ministry of Education and Research (BMBF) through the DIP project, the
James Franck German-Israel Binational Program in Laser-Matter
Interactions, and German-Israel Foundation (GIF) through grant
No. 149/2006.
\end{acknowledgments}

\end{document}